\documentclass[aps,prl,twocolumn,groupedaddress,superscriptaddress]{revtex4-1}
\usepackage{graphicx}
\usepackage{amssymb}
\usepackage{epstopdf}
\usepackage{amsmath}
\usepackage{longtable}
\usepackage{amsmath}
\usepackage{amsfonts}
\usepackage[colorlinks=true,citecolor=blue,linkcolor=red]{hyperref}
\usepackage{graphicx}
\usepackage{bbold}     
\usepackage[dvipsnames]{xcolor}

\usepackage{array,booktabs,ragged2e}
\usepackage{soul}      
\usepackage{cancel}      
\usepackage{tabularx}     
\usepackage{multirow}
\usepackage{hhline}
\usepackage{titlesec}

\begin{document}

\title{Large magnetoresistance and non-zero Berry phase in the nodal-line semimetal MoO$_2$ }

\author{Qin Chen}
\affiliation{Department of Physics, Zhejiang University, Hangzhou 310027, China}
\author{Zhefeng Lou}
\affiliation{Department of Physics, Zhejiang University, Hangzhou 310027, China}
\author{ShengNan Zhang}
\affiliation{Institute of Physics, \'{E}cole Polytechnique F\'{e}d\'{e}rale de Lausanne (EPFL), CH-1015 Lausanne, Switzerland}
\affiliation{National Centre for Computational Design and Discovery of Novel Materials MARVEL, \'{E}cole Polytechnique F\'{e}d\'{e}rale de Lausanne (EPFL), CH-1015 Lausanne, Switzerland}

\author{Binjie Xu}
\affiliation{Department of Physics, Zhejiang University, Hangzhou 310027, China}
\author{Yuxing Zhou}
\affiliation{Department of Physics, Zhejiang University, Hangzhou 310027, China}
\author{Huancheng Chen}
\affiliation{Department of Physics, Zhejiang University, Hangzhou 310027, China}
\author{Shuijin Chen}
\affiliation{Department of Physics, Zhejiang University, Hangzhou 310027, China}
\author{Jianhua Du}
\affiliation{Department of Applied Physics, China Jiliang University, Hangzhou $310018$, China}
\author{Hangdong Wang}
\affiliation{Department of Physics, Hangzhou Normal University, Hangzhou 310036, China}

\author{Jinhu Yang}
\affiliation{Department of Physics, Hangzhou Normal University, Hangzhou 310036, China}
\author{QuanSheng Wu}
\affiliation{Institute of Physics, \'{E}cole Polytechnique F\'{e}d\'{e}rale de Lausanne (EPFL), CH-1015 Lausanne, Switzerland}
\affiliation{National Centre for Computational Design and Discovery of Novel Materials MARVEL, \'{E}cole Polytechnique F\'{e}d\'{e}rale de Lausanne (EPFL), CH-1015 Lausanne, Switzerland}
\author{Oleg V. Yazyev}
\affiliation{Institute of Physics, \'{E}cole Polytechnique F\'{e}d\'{e}rale de Lausanne (EPFL), CH-1015 Lausanne, Switzerland}
\affiliation{National Centre for Computational Design and Discovery of Novel Materials MARVEL, \'{E}cole Polytechnique F\'{e}d\'{e}rale de Lausanne (EPFL), CH-1015 Lausanne, Switzerland}
\author{Minghu Fang}\email{Corresponding author: mhfang@zju.edu.cn}
\affiliation{Department of Physics, Zhejiang University, Hangzhou 310027, China}
\affiliation{Collaborative Innovation Center of Advanced Microstructure, Nanjing University, Nanjing 210093, China}
\date{\today}

\begin{abstract}
We performed calculations of the electronic band structure and the Fermi surface as well as measured the longitudinal resistivity  $\rho$$_{xx}$(\emph{T},\emph{H}), Hall resistivity $\rho$$_{xy}$(\emph{T},\emph{H}) and  quantum oscillations of the magnetization as a function of temperature at various magnetic fields for MoO$_2$  with monoclinic crystal structure. The band structure calculations show that MoO$_2$ is a nodal-line semimetal when spin-orbit coupling is ignored. It was found that a large magnetoresistance reaching 5.03 $\times$ 10$^4$$\%$  at 2~K and 9~T, its nearly quadratic field dependence and a field-induced up-turn behavior of $\rho$$_{xx}$(\emph{T}), the characteristics common for many topologically non-trivial as well as trivial semimetals, emerge also in MoO$_2$. The observed properties are attributed to a perfect charge-carrier compensation, evidenced by both calculations relying on the Fermi surface topology and the Hall resistivity measurements. Both the observation of negative magnetoresistance for magnetic field along the current direction and the non-zero Berry phase in de Haas-van Alphen measurements indicate that pairs of Weyl points appear in MoO$_2$, which may be due to the crystal symmetry breaking.
These results highlight MoO$_2$ as a new platform materials for studying the topological properties of oxides.

\end{abstract}
\pacs{}
\maketitle
\section{I. INTRODUCTION}
Since the discovery of first topological insulators fifteen years ago \cite{kane2005cl,bernevig2006ba,fu2009fu}, the important role of the topology of electronic bands in condensed matter materials has been broadly recognized. In addition to topological insulators, semimetallic topological phases with a variety quasiparticles related to the Dirac \cite{liu2014discovery,wang2013three}, Weyl \cite{shekhar2015extremely,lv2015experimental}, nodal-line \cite{an2019chiral,hu2016evidence,emmanouilidou2017magnetotransport}, nodal-chain \cite{bzduvsek2016nodal,yu2017nodal}, nodal-link \cite{yan2017nodal,chen2017topological,ezawa2017topological} band degeneracies, have been proposed and confirmed experimentally in a few dozens of materials. Topological electronic materials, as new quantum states of matter, attracted a lot of attention in the past ten years due to their promise for the development of devices with new functionalities. The search for novel topological phases and materials has become one of the most dynamic active directions of research in condensed matter physics and materials science. Recently, a number of groups  \cite{vergniory2018high,TopoMat,zhang2019catalogue,tang2019comprehensive}
predicted thousands candidate topological materials by performing systematic high-throughput computational screening across the databases of known materials. However, the topological nature of most of these candidate topological materials yet has to be confirmed by experiments. According to the predictions, many conventional materials including those currently used in electronic devices, sensors, and as energy materials, also  host topological phases. It thus becomes important to re-evaluate the physical properties of these materials from the topological view in anticipation of new technological applications.

MoO$_2$ has been widely investigated as a promising catalyst \cite{zhao2013synthesis} for electro-catalytic or photocatalytic hydrogen evolution in aqueous solution, as an efficient electrode material for lithium ion batteries \cite{zhou2011interconnected} and as a materials for supercapacitors \cite{zhang2017enhanced}. On the other hand, in order to uncover the mechanism of the metal-insulator transition in early transition metal oxides such as VO$_2$ \cite{morin1959oxides}, a well-known phenomenon originating from strong electronic correlations, both the crystal and electronic structure of MoO$_2$ have been studied \cite{moosburger2009fermi,eyert2000embedded}. Compared to the $3d$ electrons in VO$_2$, the reduced localization of the $4d$ electrons in MoO$_2$ results in metallic character with a weak electronic correlations. Recently, Zhang \emph{et} \emph{al}. \cite{zhang2019catalogue} pointed out that MoO$_2$ is a nodal-line semimetal when considering spin-orbital coupling (SOC) is neglected.

In this paper, we present calculations of the electronic band structure and the Fermi surface (FS) as well as measured the longitudinal resistivity  $\rho$$_{xx}$(\emph{T},\emph{H}), Hall resistivity $\rho$$_{xy}$(\emph{T},\emph{H}) and  quantum oscillations of the magnetization as a function of temperature at various magnetic fields for MoO$_2$  with monoclinic crystal structure.
We further reveal a nearly quadratic field dependence of magnetoresistance (MR) reaching a large value of 5.03 $\times$ 10$^4$$\%$ at 2~K and 9~T, as well as a field-induced up-turn behavior of $\rho$$_{xx}$(\emph{T}) in this material, the characteristics common for many other topologically non-trivial as well as trivial semimetals. The observed properties are attributed to a perfect charge-carrier compensation, evidenced by both calculations relying on the Fermi surface topology and the Hall resistivity measurements. Both the observation of negative MR for magnetic field along the current direction and the non-zero Berry phase in de Haas-van Alphen (dHvA) measurements indicate that pairs of Weyl points appear in MoO$_2$, which may be due to the crystal symmetry breaking.

\section{II. EXPERIMENTAL METHODS AND CALCULATIONS}

\begin{figure*}[!htbp]
\centering
\includegraphics[width=16cm]{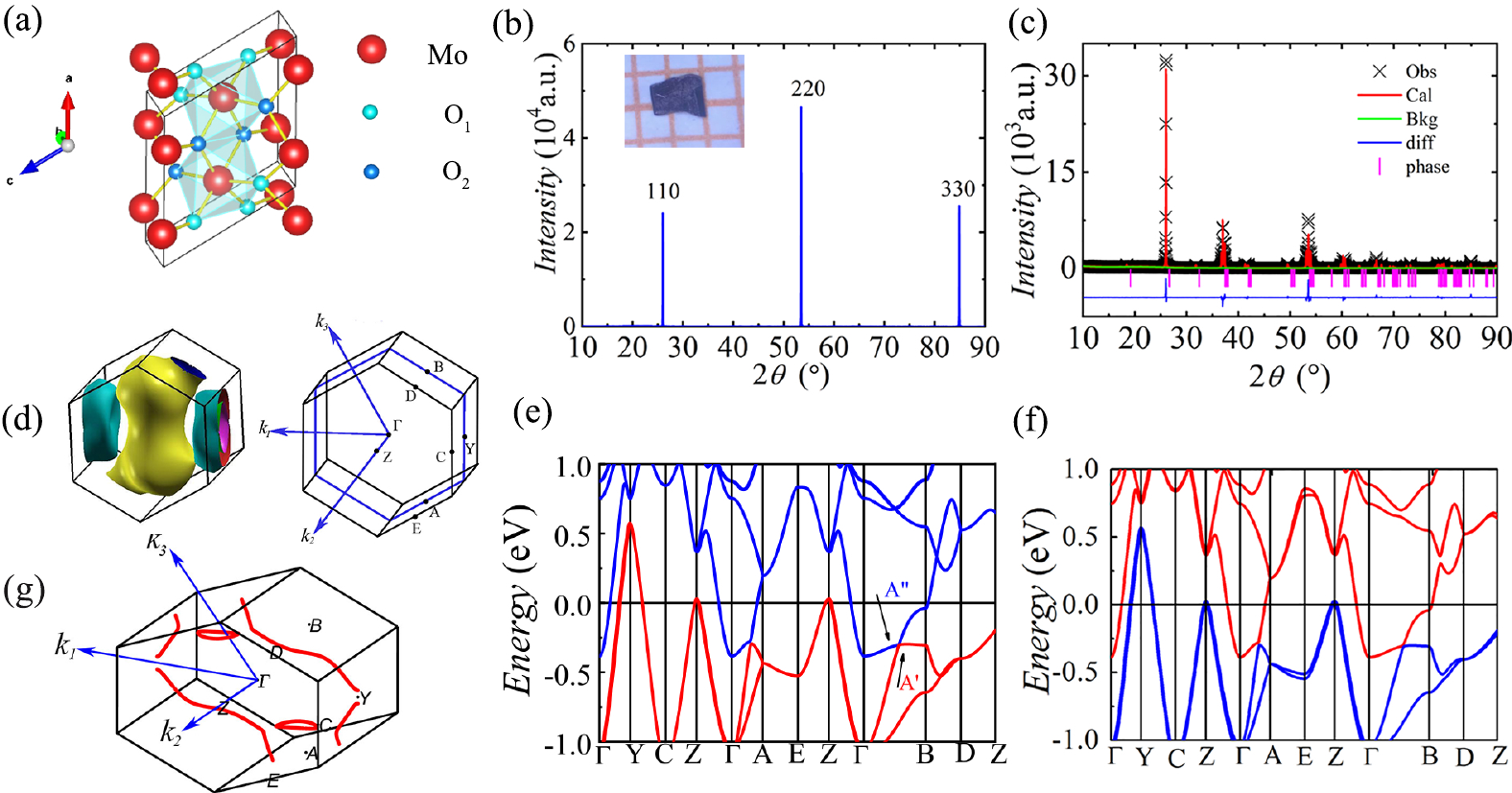}
\caption{(a) Crystal structure of monoclinic MoO$_2$. (b) Single-crystal XRD pattern. (c) XRD pattern of powder obtained by grinding MoO$_2$ crystals, the line shows its Rietveld refinement. (d) The Brillouin zone and the calculated Fermi surface of MoO$_2$. (e) and (f) Band structures of MoO$_2$ calculated without and with SOC. Valence and conduction bands are distinguished by color. (g) Nodal lines and rings with high-symmetry points in the momentum space. }
 \end{figure*}

MoO$_2$ single crystals were grown by a chemical vapor transport method. Polycrystalline MoO$_2$ prepared previously was sealed in an evacuated quartz tube with 10 mg/cm$^3$ TeCl$_4$ as a transport agent, then heated for two weeks at 1020 K, in a tube furnace with a temperature gradient of 50 K. Grey crystals with typical dimensions 1.5 $\times$ 1.0 $\times$ 0.2 mm$^3$ and a (110) easy cleavage plane [see Fig. 1(b)] were obtained at the cold end of the tube. The composition was confirmed to be Mo : O = 1 : 2 by using the energy dispersive x-ray spectrometer (EDXS). The crystal structure was determined by a powder x-ray diffractometer (XRD, Rigaku Gemini A Ultra), created by grinding pieces of crystals [see Fig. 1(c)]. It was confirmed that MoO$_2$ crystallizes in a monoclinic structure (space group \emph{P}21/c, No. 14). The lattice parameters, \emph{a} = 5.618(2) $\rm {\AA}$, \emph{b} = 4.867(1) $\rm {\AA}$ and \emph{c} = 5.637(2) $\rm {\AA}$ are obtained by using Rietveld refinement to XRD data (weighted profile factor R$_{wp}$ = 8.62$\%$, and the goodness-of-fit $\chi$$^2$ = 3.466), as shown in Fig. 1(c). Longitudinal electrical resistivity ($\rho$$_{xx}$), Hall resistivity ($\rho$$_{xy}$) and magnetization measurements were carried out by using a Quantum Design Physical Property Measurement System (PPMS, 9 T) or Magnetic Property  Measurement System (MPMS, 7 T). The band structure was calculated using the Vienna \textit{ab} \textit{initio} simulation package (VASP) \cite{kresse1996efficient,kresse1999g} with generalized gradient approximation (GGA) of Perdew, Burke and Ernzerhof (PBE) \cite{perdew1996phys} for the exchange-correlation potential. A cutoff energy of 520 eV and a $13\times15\times13$ k-point mesh were used to perform the bulk calculations.

\section{III. RESULTS AND DISCUSSIONS}
As a starting point, we discuss the results of our electronic structure calculations that provide a broad overview of MoO$_2$ and extend the initial finding of the nodal-line semimetal phase in this material~\cite{zhang2019catalogue}. In order to address the topological character of MoO$_2$, we calculate its band structure and the Fermi surface (FS). As shown in Fig. 1(d), there are three different FS sheets: two hole-like surfaces near the Y point, which occupy 9$\%$ and 2$\%$ of the Brillouin zone volume (sheets 1 and 2), respectively, and a peanut-shaped electron-like surface centered around the $\Gamma$ point that  occupies 11$\%$ of the Brillouin zone volume (sheet 3). The latter FS sheet has two loopholes near the B points, whereas it is closed along the $\Gamma$-A direction. The same volume (11$\%$) of the first Brillouin zone occupied by both electron-like and hole-like pockets indicates that monoclinic MoO$_2$ is a compensated semimetal. All FS sheets have been correctly identified using angle-resolved photoemission spectroscopy (ARPES), dHvA measurements and electronic structure calculations \cite{moosburger2009fermi}. We then identified the nodal-line band degeneracies in the band structure of this material using the open-source code WannierTools \cite{wu2018wanniertools}, which is based on the Wannier tight-binding model constructed using Wannier90 \cite{mostofi2014updated}. In the absence of SOC, the band degeneracies in MoO$_2$ are nodal lines and rings rather than point degeneracies. In other words, the band crossings exist along certain closed loops in the three-dimensional momentum space. In particular, two circular nodal rings and four nodal lines in the $\Gamma$-B-Y plane are formed.
All of the nodal lines and rings are formed by the electron and hole pockets connected by nodal points near the Fermi energy. We then use irreducible representations of the bands to analyze the nodal lines \cite{wu2018mgta}. As shown in Fig. 1(e), two bands touch at (0, 0.146, 0.246) along the $\Gamma-B$ path in momentum space.
At this touching point, the irreducible representations of two bands are $\rm{A}'$ and $\rm{A}''$ which are two different irreducible representations of point group \emph{C}$_{s}$\cite{altmann1994point}. According to Schur$^{,}$s lemma, the two bases belonging to two different irreducible representations are orthogonal to each other, hence the bands cross without gap when SOC is ignored, resulting in a nodal-line semimetal phase. Whether the topological phase transition into the Weyl phase upon introducing SOC occurs due to some crystal symmetry breaking, e.g. resulting from oxygen vacancies, element substitution or applied pressure remains to be addressed theoretically.

\begin{figure}[!htbp]
\centering
  \includegraphics[width=8cm]{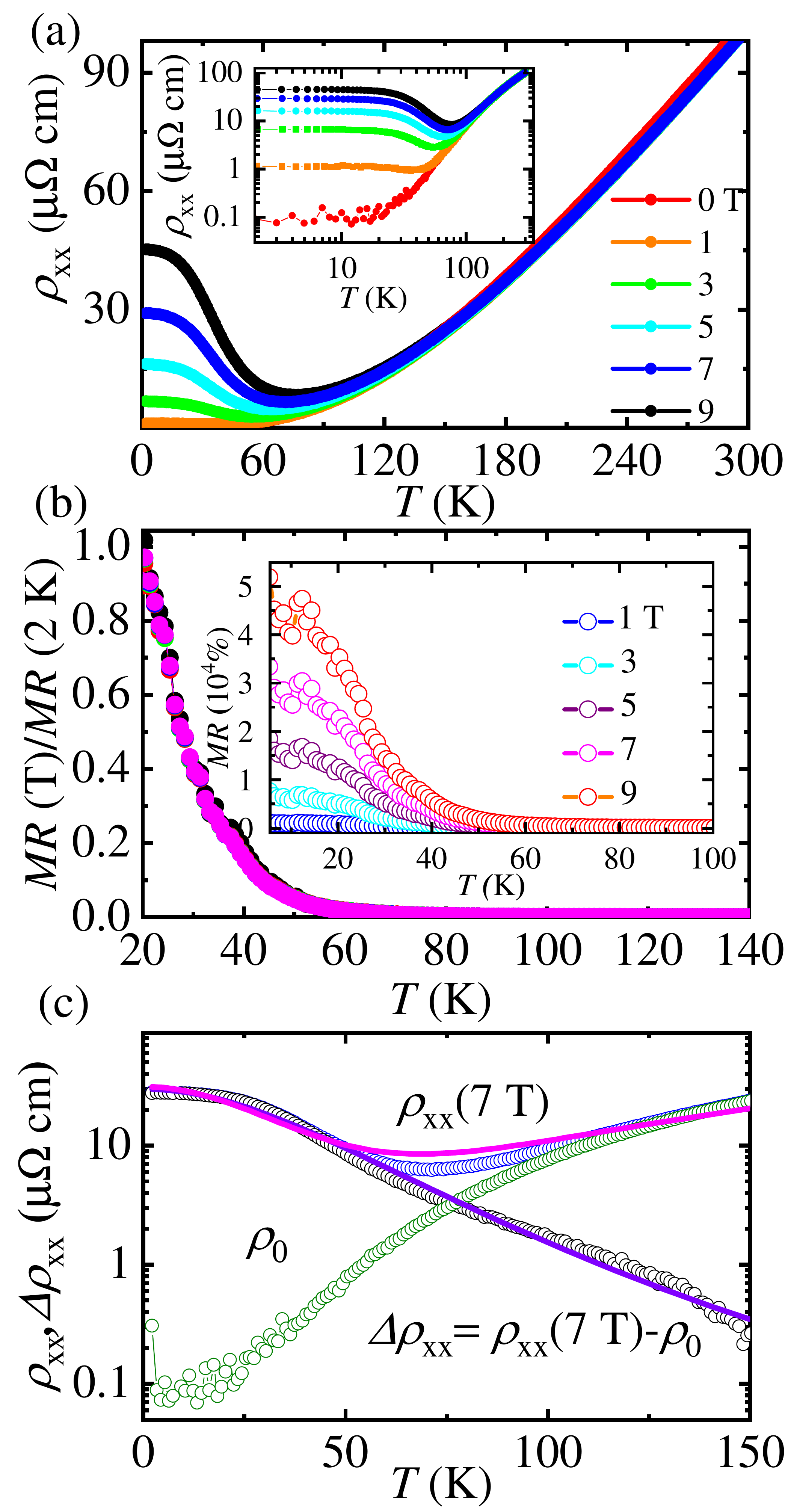}
 \caption{(a) Temperature dependence of longitudinal resistivity $\rho$$_{xx}$(\emph{T}) measured at various magnetic fields for sample 1 (S1) of MoO$_2$. (b) Temperature dependence of MR normalized by its value at 2 K at various magnetic fields. The inset shows the MR data as a function of temperature. (c) Temperature dependence of resistivity at 0 and 7 T and their difference. The solid lines are the fits of Kohler law with $\alpha$ = 0.14~($\mu$$\Omega$ cm/T)$^{1.8}$ and $m = 1.8$. }
  \end{figure}

  \begin{figure}[!htbp]
\centering
  \includegraphics[width=8cm]{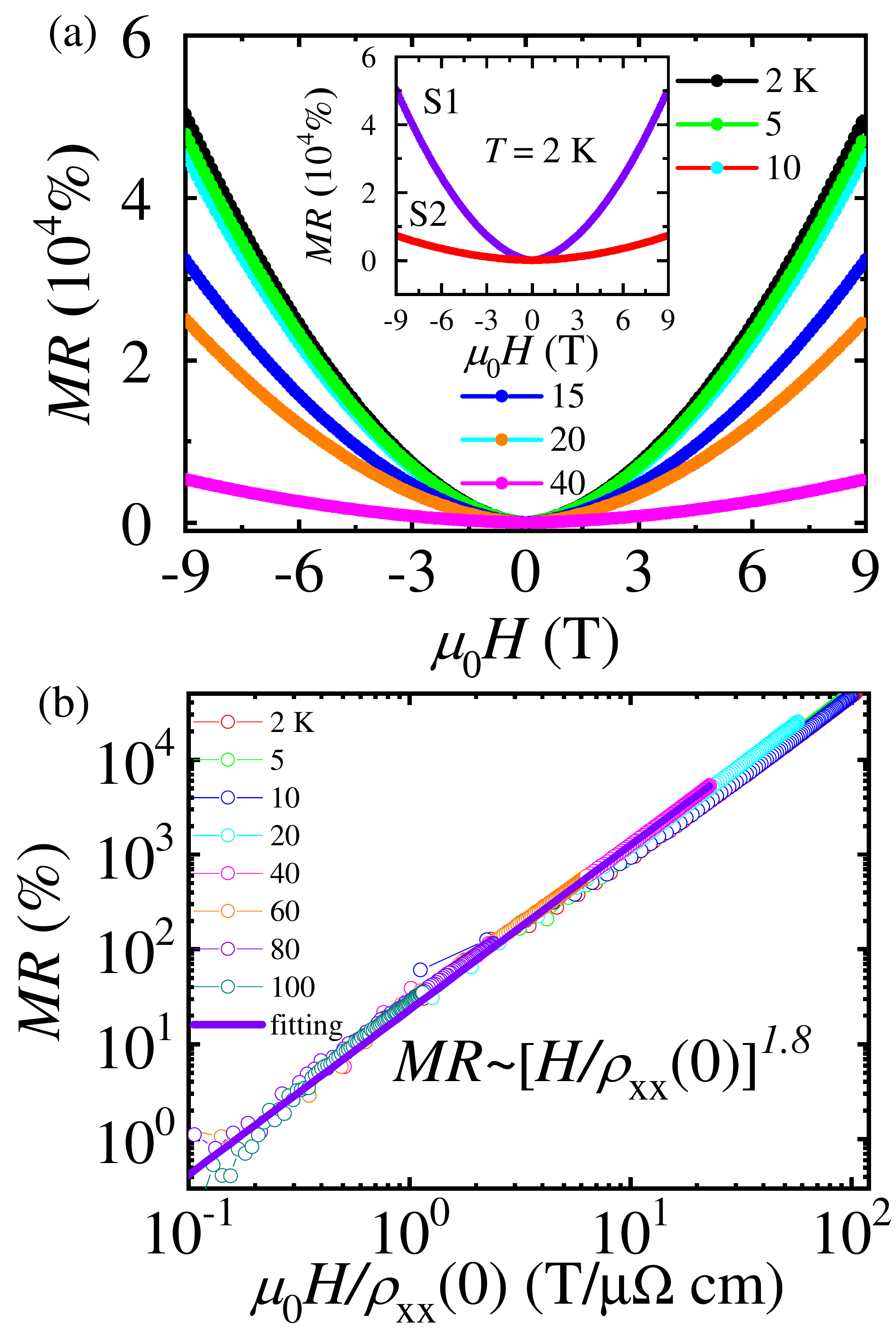}
 \caption{(a) Field dependence of MR. The inset shows the comparison of MR measured at 2~K for  sample 1 (S1) and sample 2 (S2). (b) MR plotted as a log scale as a function of \emph{H}/$\rho$$_{xx}$(0). }
  \end{figure}

  \begin{figure}[!htbp]
\centering
  \includegraphics[width=8cm]{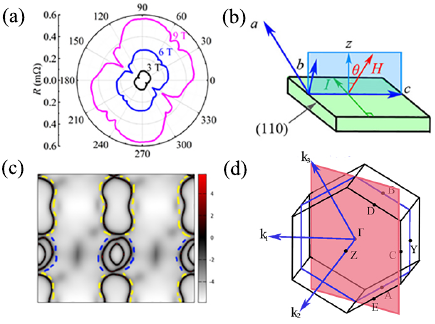}
 \caption{(a) The angular plot of resistivity at 2 K measured under various fields. (b) Scheme of applied current and magnetic field directions. Current is applied perpendicular to the \emph{c} axis, $\theta$ is the angle between \emph{z} axis and \emph{H }. (c) Cross-section of the FS of MoO$_2$ passing through the $\Gamma$ point perpendicular to the current direction. The yellow and blue dashed lines correspond to electron and hole pockets, respectively. (d)  Plane in momentum space resulting in the cross-section shown in panel (c).  }
  \end{figure}

  \begin{figure}[!htbp]
\centering
  \includegraphics[width=8cm]{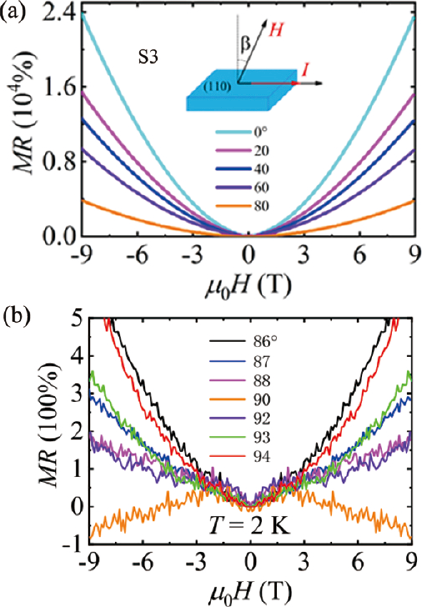}
 \caption{MR as a function of magnetic field measured at 2 K for various orientations of \emph{H} with respect to \emph{I}: (a) $\beta$ = 0$-$80$^\circ$, (b) $\beta$ = 86$-$94$^\circ$.}
  \end{figure}

Next, we focus on resistivity $\rho$$_{xx}$  measured for MoO$_2$ sample 1 (S1) at various temperatures ($\emph{T}$) and in different magnetic fields ($\mu_0$\emph{H}) with current \emph{I} $\perp$ \emph{c} axis, and $\emph{H}$ $\perp $ (110) plane. As shown in Fig. 2(a), resistivity measured at $\mu_0$\emph{H} = 0 T exhibits a metallic behavior with $\rho$(2 K) = 0.09 $\mu$$\Omega$~cm, and $\rho$(300 K) = 96 $\mu$$\Omega$~cm, thus the residual resistivity ratio (RRR) $\rho$(300 K)/$\rho$(2 K) $\sim$ 1067, indicating that this MoO$_2$ crystal (S1) has a high quality. Due to the values of resistance at low temperature reaching the measurement limitation of PPMS, $\rho$$_{xx}$(T) below 20 K is not smooth. Similar to many topological and trivial  topological semimetals \cite{takatsu2013extremely,mun2012magnetic,yuan2016large,huang2015observation,ali2014large,chen2016extremely}, MoO$_2$ exhibits extremely large magnetoresistance. As shown in Fig. 2(a), an up-turn in $\rho$$_{xx}$(\emph{T}) curves under applied magnetic field occurs at low temperatures: $\rho$$_{xx}$ increases with decreasing \emph{T} and then saturates. The inset in Fig. 2(b) shows MR as a function of temperature at various magnetic fields, with the conventional definition  $\textit{MR} = \frac{\Delta\rho}{\rho(0)} = [\frac{\rho(H)-\rho(0)}{\rho(0)}]\times100\%$. The normalized MR, shown in Fig. 2(b), has the same temperature dependence for various fields, excluding the suggestion of a field-induced metal-insulator transition \cite{zhao2015anisotropic,khveshchenko2001magnetic} at low temperatures, as discussed in our work addressing the topologically trivial semimetal $\alpha$-WP$_2$ \cite{du2018extremely}, and the work of Thoutam {\it et al.} \cite{thoutam2015temperature} on the type-II Weyl semimetal WTe$_2$. Figure 2(c) displays  $\rho$$_{xx}$(\emph{T},\emph{H}) measured at 0 and 7 T, as well as the difference $\Delta$$\rho$$_{xx}$ = $\rho$$_{xx}$(\emph{T}, 7T) - $\rho$$_{xx}$(\emph{T}, 0T). The data clearly shows that the resistivity in applied magnetic field consists of two components, $\rho$$_0$(\emph{T}) and $\Delta$$\rho$$_{xx}$, with opposite temperature dependences. As discussed by us for $\alpha$-WP$_2$ \cite{du2018extremely} and by Wang {\it et al.} for WTe$_2$ \cite{wang2015origin}, the resistivity can be written as
\begin{eqnarray}
  \rho_{xx}(T,H) = \rho_{0}(T)[1+\alpha(H/\rho_{0})^m] .
  \end{eqnarray}
The second term is the magnetic-field-induced resistivity $\Delta$$\rho$$_{xx}$, which follows Kohler$^,$s rule with two constants $\alpha$ and \emph{m}. $\Delta\rho_{xx}$ is proportional to 1/$\rho_0$ (when \emph{m } = 2) and competes with the first term upon changing temperature, possibly giving rise to a minimum in $\rho$(\emph{T},\emph{H}) curves.
Figure 3(a) shows MR as a function of field at various temperatures. The measured MR is extremely large at low temperatures, reaching 5.03 $\times 10^4\%$  at 2 K and 9 T, and does not show any sign of saturation up to the highest field (9 T) applied in our measurements. The inset in Fig. 3(a) displays MR of crystals S1 and S2 with different RRR values of 1067 and 400, respectively. It is obvious that the magnitudes of MR are strongly dependent on the quality of  samples, which was also observed in Dirac semimetal PtBi$_2$ \cite{gao2017extremely}, $\beta$-WP$_2$ \cite{kumar2017extremely}, and $\alpha$-WP$_2$ \cite{du2018extremely}. As discussed above, MR can be described by the Kohler scaling law \cite{pippard1989magnetoresistance}
\begin{eqnarray}
\textit{MR} = \frac{\Delta\rho_{xx}(T,H)}{\rho_{0}(T)} = \alpha(\textit{H}/\rho_{0})^{m} .
\end{eqnarray}
As shown in Fig. 3(b), all MR data from $\emph{T}$ = 2 to 100 K collapse onto a single straight line when plotted as MR $\sim $ \emph{H}/$\rho_{0}$ curve, with $\alpha$ = 0.14 ($\mu\Omega$ cm/T)$^{1.8}$ and \emph{m} = 1.8 obtained by fitting. The nearly quadratic field dependence of MR observed for this nodal-line semimetal MoO$_2$ is attributed to the perfect electron-hole compensation, that is an effect of the Fermi surface topology~\cite{Zhang2018mr} evidenced by the FS calculations mentioned above, as well as the Hall resistivity measurements  discussed below, both are common characteristics for most topologically  non-trivial and trivial semimetals  \cite{wang2017large,wang2015origin,du2018extremely,chen2018large}.

We also measured the anisotropy of longitudinal resistance at \emph{T} = 2 K, and $\mu_0$\emph{H} = 3, 6, and 9 T with \emph{I} $\perp$ \emph{c} axis, and by rotating the \emph{H} in the \emph{z} -\emph{c} plane. Figure 4(a) shows the angular polar plot of resistance [\emph{z} is normal to (110) plane, see Fig. 4(b)], which exhibits  mirror symmetry, i.e. \emph{R}($\theta$) =  \emph{R}($\theta$+$\pi$). The resistance data exhibits maxima at $\theta$ = 90$^\circ$, and 270$^\circ$ and minima at $\theta$ = 145$^\circ$, and 325$^\circ$, consistent with the monoclinic crystal structure. The resistance anisotropy reflects the symmetry of the projected profile of the FS onto the plane perpendicular to \emph{I}, as shown in Fig. 4(c).

In Weyl semimetals, the presence of chiral Weyl node pairs is expected to lead to an unconventional negative MR induced by the Adler-Bell-Jackiw anomaly \cite{adler1969axial,bell1969pcac} when\emph{ H} is applied parallel to electric field \emph{ E}, i.e. the current direction \emph{I}. In type-II Weyl semimetals \cite{wang2016gate}, negative MR emerges along specific crystallographic directions, referred to as the planar orientation-dependent chiral anomaly, due to the Weyl points appearing at the boundary of electron and hole pockets. To search for such chiral anomaly in MoO$_2$, we measured MR at 2 K by changing angle $\beta$ between \emph{ H } and \emph{I} as shown in the inset of Fig. 5(a). Figures 5(a) and 5(b) show MR as a function of \emph{ H } for sample 3 (S3). As \emph{ H } $\parallel$ \emph{I} ($\beta$ = 90$^\circ$), a negative MR was indeed observed except for a small positive peak below 3 T, which is usually attributed to the weak anti-localization phenomenon, as observed in both the type-I and type-II Weyl semimetals. A small misalignment of \emph{H} relative to \emph{I} leads to positive MR. We found that the emergence of such a negative MR behavior is sensitive to samples particularities and the direction of \emph{I}. We believe that whether negative MR is observed or not depends on the Fermi level position and the current direction relative to Weyl points in the  MoO$_2$ crystal, similar to the observations in the type-II Weyl semimetal WTe$_2$ \cite{wang2016gate}. The emergency of Weyl phase in MoO$_2$ may be due to the breaking of crystal symmetry resulting from, for example, the presence of oxygen vacancies in the crystal, but this hypothesis needs to be addressed in the future.
\begin{figure}[!htbp]
\centering
 \includegraphics[width=8cm]{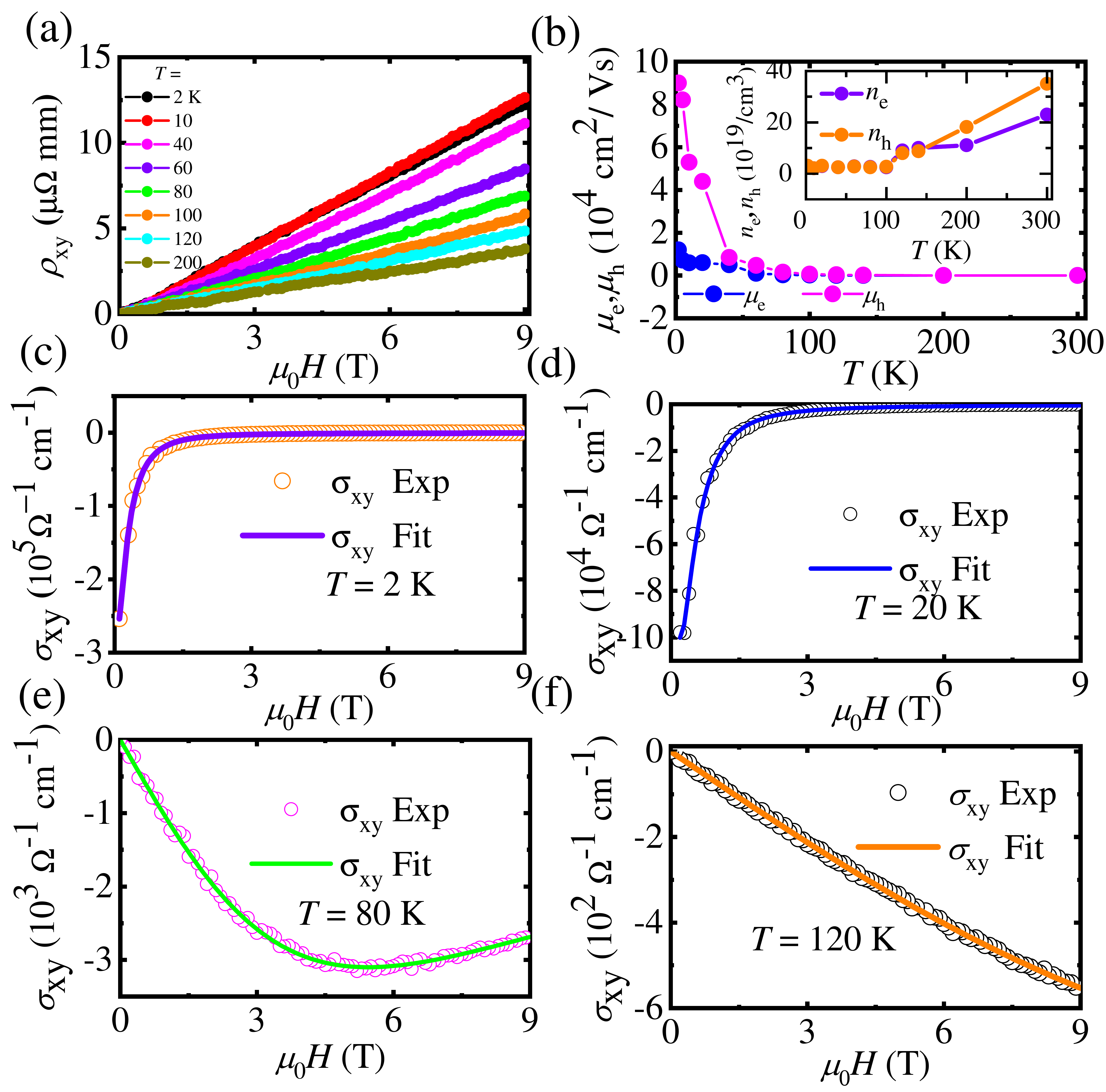}
\caption{(a) Field dependence of Hall resistivity $\rho_{xy}$ measured at different temperatures for MoO$_2$ sample 1 (S1). (b) Charge-carrier mobilities, $\mu_e$ and $\mu_h$, and (inset) the carrier concentrations, \emph{n}$_e$ and  \emph{n}$_h$, as a function of temperature. (c), (d), (e) and (f) Conductivity $\sigma_{xy}$  as a function of magnetic field at 2 K, 20 K, 80 K, 120 K. The solid lines are the fits obtained using the two-band model. }
  \end{figure}

  \begin{figure}[!htbp]
  \centering
  \includegraphics[width=8cm]{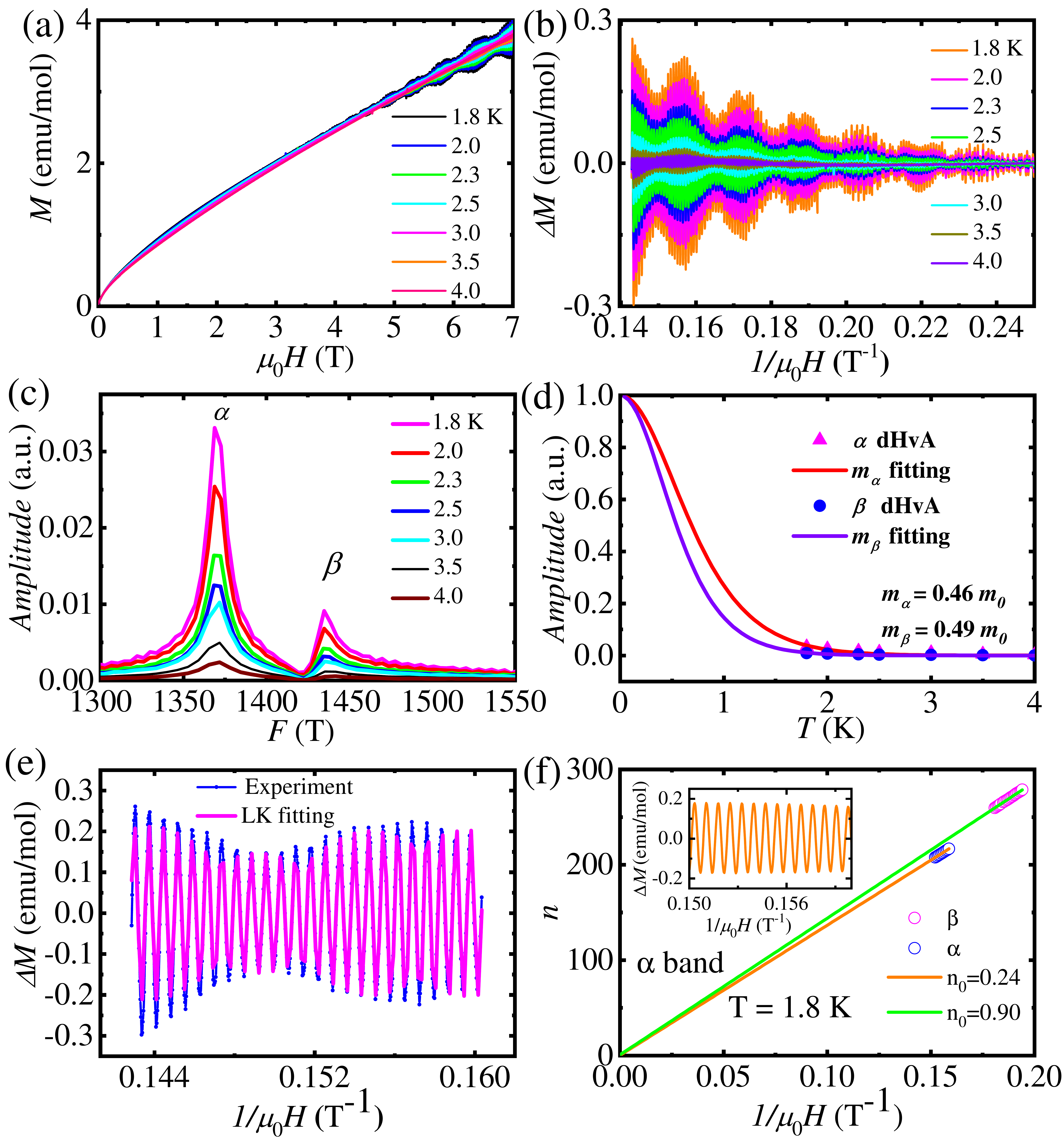}
  \caption{(a) Magnetization $M$ as a function of magnetic field measured at various temperatures. (b) dHvA oscillations plotted as a function of 1/$\mu_0$\emph{H}. (c) FFT spectra of the oscillations between 1.8 K and 4.0 K. (d) The temperature dependence of relative FFT amplitudes of each frequency and the fitting results by \emph{R}$_T$. (e) The fitting of dHvA oscillations at 1.8 K by the multi-band Lifshitz-Kosevich formula. (f) Landau-level indices fan diagram for the two filtered frequencies plotted as a function of 1/$\mu_0$\emph{H}, respectively. And the filtered wave of the $\alpha$ band (inset).}
  \end{figure}
We now discuss the characteristics of charge carriers in MoO$_2$ replying on the Hall resistivity measurements performed on sample S1. Figure 6(a) displays $\rho_{xy}$(\emph{H}) measured at various temperatures. The linear dependence of $\rho_{xy}$(\emph{H}) with a positive slope at all temperatures indicates that the holes dominate the transport properties. This is, however, not what is expected for semimetals, in which both electron and hole carriers coexist. Following the analysis of Ref. \cite{zhang2017electron,du2016large} for TaAs and TaP Weyl semimetals, we fitted the Hall conductivity tensor  $\sigma_{xy}$ = $-$ $ \rho_{xy}/(\rho_{xx}^{2}+\rho_{xy}^{2})$ using a two-carrier model \cite{hurd2012hall}
\begin{eqnarray}
\sigma_{xy} = eBn_{h}\mu_{h}^{2}/(1+\mu_{h}^{2}B^{2})-eBn_{e}\mu_{e}^{2}/(1+\mu_{e}^{2}B^{2}),
\end{eqnarray}
where \emph{ n}$_e$ and \emph{n}$_h$ denote the carrier concentrations, $\mu_e$ and $\mu_h$ denote the mobilities of electrons and holes, respectively. Figures 6(c)-6(f) display the  fits of the $\sigma_{xy}$ data measured at \emph{T} = 2 K, 20 K, 80 K and 120 K, respectively. Figure 6(b) shows the \emph{ n}$_e$, \emph{n}$_h$, $\mu_e$ and $\mu_h$  values obtained by fitting over the entire temperature range 2$-$300 K. It is remarkable that the \emph{n}$_e$ and \emph{n}$_h$ values are almost the same below 100 K, as shown in the inset of Fig. 6(b), such as at 2 K, \emph{n}$_e$ = 2.91 $\times$ 10$^{19}$ cm$^{-3}$, and \emph{n}$_h$ = 2.98  $\times $ 10$^{19}$ cm$^{-3}$. This indicates that MoO$_2$ is indeed a charge-carrier compensated semimetal, which is also consistent with the discussion of the calculated FS. It was pointed out in Ref. \cite{hu2016evidence}, that nodal-line semimetals are expected to have a higher Dirac fermion density due to the Dirac band crossings along a line or a loop compared with that in the Dirac semimetals. From the fitting of Hall resistivity using the two-band model, both electron and hole densities are estimated to be $\sim $ 10$^{19}$ cm$^{-3}$ in our sample, significantly higher than those in other Dirac semimetals such as Cd$_3$As$_2$ ($\sim $10$^{18} $ cm$^{-3}$ \cite{zhao2015anisotropic}) and Na$_3$Bi ($\sim $10$^{17} $ cm$^{-3}$ \cite{xiong2015evidence}), topological insulators ($\sim $10$^{10-12} $ cm$^{-3}$ \cite{ando2013topological}), and graphene ($\sim $10$^{10-12} $ cm$^{-3}$  \cite{zhang2005experimental}), but comparable to that of another nodal-line semimetal ZrSiS ($\sim $10$^{20} $ cm$^{-3}$  \cite{hu2016evidence}). We can reasonably expect that high carrier density in MoO$_2$ also reflects the nature of Dirac band crossings along the nodal line. Due to the existence of phonon thermal scattering at higher temperatures, as shown in Fig. 6(b), it is obvious that both the $\mu$$_e$ and $\mu$$_h$ increase notably with decreasing temperature. It is worth noting that hole mobility $\mu_h$ is almost one order of magnitude larger than $\mu_e$ at lower temperatures, e.g. at 2 K  $\mu$$_h$ = 9.0 $\times$ 10$^4 $ cm$^2$/Vs and $\mu_e$ = 1.2 $\times$ 10$^4$ cm$^2$/Vs.

Finally, the nodal-line semimetal of MoO$_2$ is supported by our measurements of the quantum oscillation of magnetization. Figure 7(a) displays the isothermal magnetization measured at various temperatures (1.8$-$4.0 K) up to 7 T ($\mu_0$\emph{H} $\parallel$ \emph{c} axis) exhibiting pronounced dHvA oscillations. In Fig. 7(b), we present the oscillation components of magnetization obtained after subtracting the background. Strong oscillations with  amplitudes of 0.2 $\sim$ 0.3 emu/mol at 1.8 K are clearly seen. From the fast Fourier transform (FFT) analyses, we obtained two frequencies 1369 T ($F_\alpha$) and 1435 T ($F_\beta$) as shown in Fig. 7(c). In general, the oscillating magnetization of 3D system can be described by the Lifshitz-Kosevich (LK) formula \cite{lifshitz1956theory,shoenberg1984magnetic} with the Berry phase \cite{mikitik1999manifestation}
\begin{eqnarray}
\Delta M \varpropto -B^{1/2}R_{T}R_{D}R_{S} \sin[2\pi(\emph{F}/\emph{B}-\gamma-\delta)] ,
  \end{eqnarray}
 \begin{eqnarray}
  R_{T}=\alpha T\mu/B \sinh(\alpha T\mu/B) ,
  \end{eqnarray}
  \begin{eqnarray}
    R_{D}=\exp(-\alpha T_{D}\mu/B) ,
  \end{eqnarray}
  \begin{eqnarray}
  R_{S}=\cos(\pi g\mu/2) ,
   \end{eqnarray}

\begin{table*}[htbp]
  \renewcommand\arraystretch{1.2}
  \centering
  \caption{The obtained parameters by fitting dHvA data for MoO$_{2}$}\label{1}
  \setlength{\tabcolsep}{7mm}
  {
  \begin{center}
  \begin{tabular}{ccccc}

    \hline
    \hline
    Parameters & F$_{\alpha}$(LK) & F$_{\beta}$(LK) &F$_{\alpha}$(Landau fan)& F$_{\beta}$(Landau fan) \\
    \hline
    Frequency (T)& 1369 & 1435 &  \\
    \textit{m}$^{*}$/\textit{m}$_{0}$ & 0.463 & 0.498  \\
    \emph{T}$_{D}$ (K)& 2.05 & 0.91 \\
    $\tau_{q}$ (ps)& 0.6 & 1.3 \\
    $\mu_{q}$ (cm$^{2}$/Vs)& 2300 & 4600 \\
    $\varphi_{B}$($\delta$=+1/8) & 0.83$\pi$ & 1.95$\pi$&0.73$\pi$ & 2.05$\pi$ \\
    $\varphi_{B}$($\delta$=$-$1/8) & 0.33$\pi$  & 1.45$\pi$ & 0.22$\pi$ & 1.55$\pi$ \\

    \botrule

  \end{tabular}
   \end{center}
  }
\end{table*}
\noindent
where $\mu$ is the ratio of effective cyclotron mass \emph{m}$^*$ to free electron mass \emph{m}$_0$, \emph{T}$_\emph{D}$ is the Dingle temperature, and $\alpha$ = ($2\pi^2 \emph{k}_\emph{B} m_0$)/($\hbar$e). The phase factor $\delta$ = 1/8 or $-$1/8 for 3D system. The effective mass \emph{m}$^*$ can be obtained by fitting the temperature dependence of the oscillation amplitude \emph{R}$_\emph{T}$(T) as shown in Fig. 7(d). For $F_\alpha$ = 1369 T and $F_\beta$ = 1435 T, the obtained \emph{m}$^*$ is  0.463 m$_0$, and 0.498 m$_0$, respectively, which is smaller than that of MoP$_2$ (1.6 m$_0$) \cite{wang2017magnetotransport}, but larger than that of the nodal-line semimetals ZrSiS (0.16 m$_0$) \cite{ali2016butterfly} and ZrSiSe (0.082 m$_0$), ZrSiTe (0.093 m$_0$) \cite{hu2016evidence}. Using the fitted \emph{m}$^*$ as a known parameter, we can further fit the oscillation patterns at given temperatures [e.g., \emph{T} = 1.8 K, see Fig. 7(e)] to the LK formula with two frequencies, from which quantum mobility and the Berry phase can be extracted. The fitted values of \emph{T}$_\emph{D}$ are of 2.05 K and 0.91 K, which corresponds to the quantum relaxation times $\tau_q$ = $\hbar$/(2$\pi\emph{$k_BT_D$}$) of 0.6 ps and 1.3 ps; and quantum mobility $\mu_q$ (e$\tau$/\emph{m}$^*$) of 2300 cm$^2$/Vs and 4600 cm$^2$/Vs for $F_\alpha$ and $F_\beta$, respectively, as listed in Table I. The LK fit also yields a phase factor $-\gamma-\delta$ of $-$0.21 ($F_\alpha$), from which the Berry phase $\phi_\emph{B}$ is determined to be 0.83 $\pi$ for $\delta$ = 1/8 and 0.33 $\pi$ for $\delta$ = $-$1/8. The phase factor for $F_\beta$ is 0.35, with $\phi_\emph{B}=1.95 \pi$ ( $\delta$ = 1/8) and 1.45 $\pi$ ($\delta$ = $-$1/8).

The Berry phase can also be obtained from the commonly used Landau level fan diagram \cite{hu2017nearly} (i.e., the LL indices \emph{n} plotted as a function of the inverse of magnetic field 1/$\emph{B}_n$). According to customary practice, the integer LL indices \emph{n} should be assigned when the Fermi level lies between two adjacent LLs \cite{xiong2012high}, where the density of state near the Fermi level DOS (\emph{E}$_F$) reaches a minimum. Given that the the oscillatory magnetic susceptibility is proportional to the oscillatory DOS at the Fermi level (i.e. $\Delta$(d\emph{M}/d\emph{B}) $\propto$ $\Delta$DOS ($\emph{E}_F$)) and that the minima of $\Delta$\emph{M} and d($\Delta$\emph{M})/d\emph{B } are shifted by $\pi$/2, the minima of $\Delta$\emph{M} should be assigned to $n-1/4$. The established LL fan diagram based on this definition is shown in Fig. 7(f). The extrapolation of the linear fit in the fan diagram yields an intercept $\emph{n}_0$ = 0.24, which appears to correspond to a Berry phase $\phi _\emph{B}$ = 2$\pi$(0.24+$\delta$) of  0.73$\pi$ ($\delta$ = 1/8) and 0.22$\pi$ ($\delta$ = $-$1/8) for the $F_\alpha$ band. This is consistent with the results of the LK formula. The non-zero Berry phase for two bands indicates that monoclinic MoO$_2$ to be a topological nodal-line semimetal.

\section{IV. CONCLUSION}
In summary, we performed electronic structure calculations and measured the longitudinal resistivity, Hall resistivity and quantum oscillations of magnetization of the monoclinic phase of MoO$_2$ single crystals. It was found that MoO$_2$ exhibits many common characteristics for topological and trivial semimetals, such as large values of MR, reaching 5.03 $\times$ 10$^4$$\%$ at 2 K and 9 T, its nearly quadratic field dependence, and a field-induced up-turn behaviour of $\rho$$_{xx}$(\emph{T}). Both the calculated Fermi surface and Hall resistivity measurements attribute these magnetotransport properties to perfect charge-carrier compensation. Interestingly, we observed an unconventional negative MR when the magnetic field was applied along the \emph{I} direction in some samples as well as the non-zero Berry phase in the dHvA measurements. These results indicate that pairs of Weyl points appear in MoO$_2$, which may be due some source of crystal symmetry breaking, e.g. resulting from the presence of oxygen vacancies. These results highlight MoO$_2$ as an intersting new material for studying the topological properties of oxides.

\section{ACKNOWLEDGEMENTS}
This research is supported by the Ministry of Science and Technology of China under Grants No. 2016YFA0300402 and No. 2015CB921004 and the National Natural Science Foundation of China (NSFC) (No. 11374261, No. 11974095), the Zhejiang Natural Science Foundation (No. LY16A040012) and the Fundamental Research Funds for the Central Universities. S.N.Z, Q.S.W. and O.V.Y. acknowledge support by the NCCR Marvel. First-principles calculations were partly performed at the Swiss National Supercomputing Centre (CSCS) under projects s1008 and mr27 and the facilities of Scientific IT and Application Support Center of EPFL.

\bibliography{document}
\end{document}